\documentclass[reviewcopy]{elsarticle}

\usepackage[reviewcopy]{adndt}
\usepackage{longtable}

\usepackage{amsmath}

\usepackage{amssymb}

\biboptions{square,sort&compress}
\bibpunct[]{[}{]}{,}{n}{}{;}
\begin{document}

\begin{frontmatter}

\journal{Atomic Data and Nuclear Data Tables}


\title{Energy levels and radiative rates for transitions in S-like Sc~VI, V VIII, Cr IX, and Mn~X}

  \author[One]{Kanti M. Aggarwal\fnref{}\corref{cor1}}
  \ead{K.Aggarwal@qub.ac.uk}



  \cortext[cor1]{Corresponding author.}

  \address[One]{Astrophysics Research Centre, School of Mathematics and Physics, Queen's University Belfast,\\Belfast BT7 1NN,
Northern Ireland, UK}


\date{16.12.2002} 

\begin{abstract}  
Energy levels, radiative rates and lifetimes are reported for four S-like ions, namely  Sc~VI, V VIII, Cr IX, and Mn~X. Two independent atomic structure codes, namely the general-purpose relativistic atomic structure package ({\sc grasp})  and the flexible atomic code ({\sc fac}),    have been adopted for calculating the energy levels, with differing amounts of configuration interaction. This is mainly to make some assessment  of accuracy. However, the {\sc grasp} alone is used for calculating the remaining parameters.  Results are reported for varying number of levels of these ions, and for calculating lifetimes contributions are included from all types of transitions, i.e. E1, E2, M1, and M2. Comparisons are made with the earlier available experimental and theoretical results and assessments of accuracy are given for each ion. Additionally, the presently reported data cover a significantly larger number of levels and transitions than already available in the literature for the four S-like ions. \\ \\

{\em Received}: 23 May 2019, {\em Accepted}: 13 June 2019

\vspace{0.5 cm}
\end{abstract}

\end{frontmatter}




\newpage

\tableofcontents
\listofDtables
\listofDfigures
\vskip5pc


\section{Introduction}

Iron group elements (Sc to Zn) are very important for  studies of astrophysical plasmas, because many of their lines are frequently observed from different ionization stages. These lines are helpful in understanding  the plasma characteristics, such as temperature, density and chemical composition.  Particularly important among these are the ions of Sc, Fe and Ni. The latter two have rich spectrum in a variety of astrophysical plasmas -- see for example the {\em Atomic Line List} (volume 2.04) of Peter van Hoof at {\tt www.pa.uky.edu/{$\sim$}peter/atomic/} or the CHIANTI website at   {\tt  www.chiantidatabase.org}, whereas Sc is a key element in (particularly) understanding the Am stars \cite{am}. This element also belongs to the group of rare earths \cite{kma1} and hence is equally important for the studies of fusion plasmas, as indeed the others are,  because these are often impurities  in walls of fusion reactors. Therefore, for a variety of reasons, atomic data (including energy levels and oscillator strengths or radiative decay rates) are required for many ions. The need for atomic data has significantly increased  with the developing ITER (international thermonuclear experimental reactor) project. Generally, some atomic data (experimental as well as theoretical) for all these elements and their ions are available in the literature, but these are not (fully) sufficient for the diagnostics and/or the modelling of plasmas. Experimental data are mostly limited to a {\em few}\, energy levels, and have been compiled and assessed  by the NIST (National Institute of Standards and Technology) team, and their recommendations are freely available at their website {\tt http://www.nist.gov/pml/data/asd.cfm}.  We discuss below the available theoretical data and the necessity for improvements and/or extension. 

Since it is not possible to discuss all ions of the iron group elements in a single paper, we confine ourselves to S-like ions of Sc, V, Cr, and Mn, i.e Sc~VI, V VIII, Cr IX, and Mn~X. It may be noted that Ti~VII has been deliberately omitted because atomic data for it have already been reported \cite{tivii}. There are a few calculations available in the literature for these ions, see the NIST website for references,  but the most comprehensive and accurate results are those by  Froese Fischer et al. \cite{mchf}. For the calculations they have adopted their multi-configuration Hartree-Fock (MCHF) method,  have included very large configuration interaction (CI), and have reported energies for about 40 levels, belonging to the  3s$^2$3p$^4$, 3s3p$^5$ and  3s$^2$3p$^3$3d  configurations, which generate  47 levels  in total. However, these levels are {\em not} the lowest in energy, because several from the other configurations, such as 3p$^6$ and 3s3p$^4$3d, intermix {\em and} they have ignored levels of  angular momentum $J >$ 3. Similarly, they have neither reported oscillator strengths (f-values) or radiative rates (A-values) for all transitions among the calculated levels nor for the higher types, i.e. magnetic dipole (M1), electric quadrupole (E2) and magnetic quadrupole (M2). These results are also useful, apart from the dominant and most important E1 (electric dipole) type,  for the further calculations of lifetimes as well as for plasma modelling. Therefore, there is a clear scope for extension of their data.

Realising the importance of Sc ions and considering the paucity of atomic data for these, Massacrier and Artru  \cite{ma} performed calculations for several ions of this element, i.e. Sc~III to Sc~XXI. For S-like Sc~VI, they considered 1889 levels which included 51  configurations with up to $n$ = 10. For the calculations they adopted the {\em flexible atomic code}\,(FAC) of Gu \cite{fac}, currently available at the website {\tt https://www-amdis.iaea.org/FAC/}. In comparison to the calculations of Froese Fischer et al. \cite{mchf}, the one by Massacrier and Artru included much less CI, and hence there is scope for improvement, because Sc is only moderately heavy. Additionally, level designations is a big problem in the presentation of energy levels and straightforward calculations, such as by Massacrier and Artru, do not distinguish the levels from different combinations of the orbitals, and as a result the same designation is given to several levels, which makes it hard(er) for applications and comparisons. For example, the configuration 3p$^6$ generates only one level (i.e. $^1$S$_0$), but has been listed twice by them -- see  levels 47 and 164 in  their table~5. Although level designations cannot be uniquely resolved for all levels in a large calculation, obvious discrepancies like the one noted here can definitely be avoided. We will discuss about it more in later part of the paper.

In two recent papers \cite{elm1, elm2}, El-Maaref et al. have reported atomic data for two S-like ions, namely Sc~VI and Mn~X. Compared to the work of Froese Fischer et al. \cite{mchf}, they have reported data for a larger number of levels, i.e. 161 of the 3s$^2$3p$^4$, 3s3p$^5$, 3s$^2$3p$^3$3d, 3s$^2$3p$^3$4s/4p/4d, and 3s$^2$3p$^3$5s/5p configurations, i.e. 8 in total. Furthermore, to have confidence in the results for accuracy and reliability, they have performed calculations with two different codes, namely the {\em configuration interaction version 3} (CIV3: \cite{civ3}) and the Los Alamos National Laboratory (LANL) code, which is mainly based on the Hartree-Fock relativistic code of Cowan \cite{cow}. Unfortunately, in spite of these calculations being more recent and hence `expected' to be at least as  accurate (if not more) than the earlier ones, are highly deficient, inaccurate and unreliable for several reasons, as has been highlighted and explained by us \cite{sc6, mn10}. Furthermore, not only have they  considered very limited CI in the calculations, the choice of configurations is arbitrary, because only those have been included for which results are available on the NIST website. As a result of this their reported levels are not the lowest in energy, and this seriously affects the further calculations of lifetimes ($\tau$), because contributions from the {\em missing} levels are not accounted for. Although we will elaborate on this more in later sections, the requirement for a larger, but accurate and reliable, data remains.
 
In our work,  we employ the general-purpose relativistic atomic structure package ({\sc grasp}), which is a fully relativistic code and is based on the multi-configuration Dirac-Fock (MCDF) method. The  original version (referred to as GRASP0) was first published by Grant et al. \cite{grasp0}, but has undergone through several revisions and modifications by many workers, and the one adopted here has been  modified by (one of the authors) P.~H.~Norrington, and is presently hosted at the website  {\tt http://amdpp.phys.strath.ac.uk/UK\_APAP/codes.html}.  Compared to our earlier work on Ti~VII \cite{tivii}, we have considered a larger CI for the ions of present interest, but it still has its limitations because some of the configurations generate a very large number of levels. Therefore, to assess the accuracy of our calculated energies as well as the significance of additional CI, we have also  performed calculations  with FAC, which is not only as relativistic as GRASP is but is also very efficient in running. Furthermore, the main advantage of this code is that it can handle very large CI and generally produces data of comparable (and acceptable) accuracy, although giving specific designations to all levels is a greater problem in this.

\section{Energy levels}

The GRASP code has several options for the optimisation of orbitals, but we have used (our preferred)  choice of  `extended average level' (EAL),  in which a weighted (proportional to 2$j$+1) trace of the Hamiltonian matrix is minimised.  The contributions of Breit and quantum electrodynamic effects (QED) are also included in the code, although these are comparatively more  significant for the heavier ions.   Following our experience with Ti~VII \cite{tivii},  some further tests with a number of configurations, and keeping our computational resources (and the limitations of the code) in mind, our final calculations include 4498 levels of 41 (10 more than for Ti~VII)  configurations, which are: 3s$^2$3p$^4$, 3s3p$^5$, 3s$^2$3p$^3$3d, 3s$^2$3p3d$^3$, 3p$ ^6$, 3s3p$^4$3d, 3p$^5$3d, 3s$^2$3p$^2$3d$^2$, 3s3p$^3$3d$^2$,  3s3p$^2$3d$^3$,  3p$^4$3d$^2$, 3s$^2$3p$^3$4$\ell$, 3s3p$^4$4$\ell$, 3s$^2$3p$^2$3d4$\ell$,  3s3p$^3$3d4$\ell$, 3p$^5$4$\ell$, 3s$^2$3p$^3$5$\ell$, and 3s$^2$3p$^3$6$\ell$ ($\ell \le$ g). The highest energy range covered by these levels is up to $\sim$21 Ryd.  Since CI for the ions under consideration here is  very important, we have therefore performed a series of calculations with FAC, by gradually increasing the CI, and our {\em final} one includes 1~59~162 levels (or configuration state functions, CSF) arising from all possible combinations of the 3*6, 3*5 4*1, 3*5 5*1, and 3*5 6*1  configurations, i.e. 242 in total. These large calculations are similar to our earlier work on Ti~VII, and are sufficient for assessing the accuracy of energy levels as well as the impact of additional CI, if any. In Tables~1-4 we list our calculated energies from both codes (GRASP and FAC), along with the compilations of NIST and other available results, for Sc~VI, V VIII, Cr IX, and Mn~X, respectively, and below we discuss our results in detail.

\subsection{Sc~VI}

In Table~1 are listed energies for (the lowest) 61 levels of Sc~VI, which mostly belong to the  3s$^2$2p$^4$, 3s2p$^5$, 3s$^2$3p$^3$3d,  and 3p$^6$ configurations. Beyond these  there is a mixing with those from others, particularly of 3s$^2$2p$^3$4$\ell$. However, results for higher levels can be obtained from the author on request. Included in the table are our results from GRASP and FAC, those of Froese Fischer et al. \cite{mchf} from MCHF and El-Maaref et al. \cite{elm1} from CIV3, and the recommendations of NIST. For the levels listed here none of the earlier results are available for all of them, as there are significant gaps in between, particularly beyond 45. It may also be noted that the energies for some of the 3p$^3$($^2$P)3d and 3p$^3$($^2$D)3d levels of NIST have been swapped, because in our opinion their listings are not correct. This conclusion is based on (i) comparisons with other theoretical energies as well as (ii) a closer look at the mixing coefficients from different levels in our calculations. For example, levels 26 and 45 are 3p$^3$($^2$P)3d~$^1$D$^o_2$ and 3p$^3$($^2$D)3d~$^1$D$^o_2$ in all theoretical works, but are listed in the reverse order in the NIST compilation. In our calculations with GRASP their respective mixing coefficients are 0.866~(3p$^3$($^2$P)3d)~$^1$D$^o_2$ $-$ 0.432~(3p$^3$($^2$D)3d)~$^1$D$^o_2$ and 0.863~(3p$^3$($^2$D)3d)~$^1$D$^o_2$ + 0.424~(3p$^3$($^2$P)3d)~$^1$D$^o_2$. Thus there is no ambiguity in level designations and, more importantly, all our calculations with varying amount of CI provide the similar mixing and hence the designations. Therefore, we are confident of our listings. We will also like to note here that for two levels 18 ($^1$S$_0$) and 41 ($^1$P$^o_1$) there is a small typographical error in the listings of Froese Fischer et al.  (see their table~6), because they have designated these as $^1$D$_0$ and $^1$D$^o_1$, respectively. Similarly, energies of El-Maaref et al. are misplaced for two levels, namely 18 ($^1$S$_0$) and 37 ($^3$S$^o_1$).

Generally, there are no significant discrepancies between our energies with GRASP and those of NIST, as the differences are within 0.1~Ryd, except for two levels, 45 (3p$^3$($^2$D)3d~$^1$D$^o_2$ ) and 47 (3p$^3$($^2$D)3d~$^1$F$^o_3$), for which our energies are (slightly) higher by 0.13~Ryd. For both these  levels the energies of Froese Fischer et al. \cite{mchf}  and El-Maaref et al. \cite{elm1} match better with those of NIST and hence indicate a scope for improvement in our calculations. Indeed, our larger calculations with FAC are closer to those of NIST (and MCHF) as the differences have reduced by a factor of two. For a few levels  the FAC energies are lower than with GRASP by up to a maximum of 0.05~Ryd, and are assessed to be comparatively more accurate. To conclude, we may state that there are no large discrepancies in energies for the levels of Sc~VI, but the earlier available theoretical or experimental results have gaps in their listings, which we have addressed in the present work. 

\subsection{V VIII and Cr IX}

In Tables~2-3 we list our energies from GRASP and FAC for the lowest 110 and 143 levels of V~VIII and Cr ~IX, respectively. The criterion for choosing these number of levels is the same as for Sc~VI, i.e. the cut off is implemented just before the levels of the 3s$^2$2p$^3$4$\ell$ configuration start appearing, but data for higher levels  can be obtained from the author on request.  For these two ions the NIST listings are for only a few levels and the only reliable theoretical energies with which to compare are of Froese Fischer et al. \cite{mchf} from the MCHF code. Both of these results are included in the tables. As for Sc~VI, for these two (and Mn~X as well) ions also Froese Fischer et al. have mis-labelled  two levels as $^1$D$_0$ and $^1$D$^o_1$ in stead of $^1$S$_0$ and $^1$P$^o_1$, and similarly, their listed energies are not for the lowest levels as there are gaps. For the common levels the differences between the NIST and GRASP energies are  insignificant, except for some higher ones for which the latter are higher by up to 0.15~Ryd. Nevertheless, the MCHF energies are comparatively closer to those of NIST, and so are ours with FAC, because of much larger CI. Overall, the comparisons and conclusions are the same as for Sc~VI, i.e. there are no appreciable discrepancies, in magnitudes or orderings, for V~VIII and Cr~IX, but the earlier results have several gaps for which no energies are available and we have filled that in. Finally, we will like to state that for these two ions labellings for all levels are (although) unique, but may be changeable for a few (4 for V~VIII and 10 for Cr~IX), because the coefficient from the same eigenvector dominates in two levels. As an example, levels 75 and 77 are (3s3p$^4$($^2$D)3d)~$^3$F$_2$ and $^1$D$_2$, respectively for both ions, but their dominant compositions are 0.54~$^3$F$_2$+0.48~$^1$D$_2$ and 0.66~$^3$F$_2$+0.43~$^1$D$_2$ for V~VIII and 0.55~$^3$F$_2$+0.47~$^1$D$_2$ and 0.66~$^3$F$_2$+0.43~$^1$D$_2$ for Cr~IX, i.e. the coefficient for $^3$F$_2$ dominates in both levels and both ions. Another calculation with a different method, code or CI may change their compositions, or the authors may prefer to interchange the designations. These problems are common in any atomic structure calculation and therefore a caution needs to be exercised while making comparisons.

\subsection{Mn X}

For this ion too, NIST energies are available for only a few levels which are listed in Table~4,  along with our results with GRASP and FAC, and those of  El-Maaref et al. \cite{elm2} from CIV3. The corresponding energies of Froese Fischer et al.~\cite{mchf} from MCHF  are not listed here because the comparisons are the same as for the other three S-like ions. Our energies are for the lowest 190 levels whereas those of El-Maaref et al. (or Froese Fischer et al.) are for only 45, although they have reported energies for some  higher levels involving the 4$\ell$ and 5$\ell$ configurations. For the lowest 9 levels there are no discrepancies between our results with GRASP and those of NIST, but for a few higher ones (41--46) our results are larger by up to 0.2~Ryd. An extensive CI, as included in the FAC calculations, has not made any significant difference, because the discrepancies are still up to 0.14~Ryd. A level which requires particular attention is 39, i.e. 3s$^2$3p$^3$($^2$D)3d~$^1$P$^o_1$, which is misplaced by NIST. In our opinion it should be 53, i.e. 3s$^2$3p$^3$($^2$P)3d~$^1$P$^o_1$, and in our calculation it is clearly identifiable with its coefficient of 0.94.   However, level 39 is well mixed as \\0.42~(3s3p$^5$)~$^1$P$^o_1$+0.30~(3s$^2$3p$^3$($^2$D)3d)~$^3$P$^o_1$+0.62~(3s$^2$3p$^3$($^2$D)3d)~$^3$S$^o_1$+0.50~(3s$^2$3p$^3$($^2$D)3d)~$^1$P$^o_1$. The highest coefficient for this level is of $^3$S$^o_1$, which is also dominant in level 38 (with a mixing coefficient of 0.72),  and we have therefore designated level 39 on the basis of the second highest contribution. However for $\sim$15 levels there may be some ambiguity in their designations (because of the heavy mixing), as noted for the other  two ions, V~VIII and Cr~IX.

In conclusions, we may confidently state that our energies from FAC for all levels are slightly more accurate than from GRASP, but there are no (major) discrepancies in level orderings, and our results cover a much larger number of levels than in the earlier works, and can therefore be reliably employed in any modelling application.

  \section{Radiative rates}\label{sec.eqs} 
 
 Our calculated results  with the GRASP code for  transition energies (wavelengths, $\lambda$ in ${\rm \AA}$), radiative rates (A-values in s$^{-1}$), oscillator strengths (f-values, dimensionless), and line strengths (S-values, in atomic units, 1 a.u. = 6.460$\times$10$^{-36}$ cm$^2$ esu$^2$) for E1 transitions are listed  in Tables 5--8 for Sc~VI, V VIII, Cr IX, and Mn~X, respectively.  These results have been obtained in two gauges, i.e.  velocity  and length or Coulomb and Babushkin, respectively, and therefore their ratio (R) has also been listed in the last columns, which gives an indication of the accuracy of the data. For strong transitions (say f $>$ 0.1) R is generally  close(r) to unity, but for the weak(er) ones with small f-values it may (sometimes) differ substantially. Additionally,  for the corresponding  E2, M1 and M2 transitions  only  the A-values are listed, because  data for f- or S-values can be easily   obtained using Eqs. (1-5) given in \cite{tivii}.   The  indices used in these tables to represent the lower and upper levels of a transition correspond to those defined in Tables 1--4 for the respective ions. Furthermore,  for brevity only transitions from the lowest 5 to higher excited levels are listed in Tables 5--8, but  complete data for all transitions, in the ASCII format,  are being made available (online) in the electronic version of the paper.
 
 As stated earlier in Section~1, some results are already available in the literature with which to make comparisons, and subsequently an assessment of accuracy. These are mainly by Froese Fischer et al. \cite{mchf} for all S-like ions under consideration here, and by El-Maaref et al. \cite{elm1, elm2} for Sc~VI and Mn~X. Therefore, in Tables A and B we make comparisons for these two ions alone, which should be sufficient for the required purpose. 
 
In Table~A we compare our f-values from GRASP with those of Froese Fischer et al. \cite{mchf}  with MCHF and  El-Maaref et al. \cite{elm1} with CIV3 as well as LANL codes for some E1 transitions of Sc~VI.  Firstly, there is no agreement between the CIV3 and LANL f-values, as differences for some are up to three orders of magnitude, see for exmple transitions 1--38 and 3--39, both of which are rather strong, and hence a much better agreement is expected. On the basis of the comparisons made in Section~2 for the energy levels, their CIV3 f-values should be comparatively more accurate, but are unfortunately worse, as the agreements with the LANL f-values are more satisfactory for most transitions listed here. Differences between the LANL and other f-values are also significant (up to a factor of four) for several transitions, such as 1--35/38/39, 2--33/34 and 3--31. Therefore, we do not assess their results to be accurate and reliable, and neither these are available for all transitions among the levels calculated by them. 

In Table~B we compare our f-values from GRASP with those of Froese Fischer et al. \cite{mchf}  with MCHF and  El-Maaref et al. \cite{elm2} with CIV3 as well as LANL codes for some E1 transitions of Mn~X. For this ion disagreements between the CIV3 and LANL f-values are not as stark as for Sc~VI seen in Table~A, and are within a factor of five -- see for example, transitions 2--40/41/43/44, 3--29/44 and 4--53. Also note that El-Maaref et al. have incorrectly listed f-values for two transitions, namely 4--26/45 (3s$^2$3p$^4$~$^1$D$_2$ -- 3s$^2$3p$^3$($^2$P)3d~$^1$D$^o_2$/3s$^2$3p$^3$($^2$D)3d~$^1$D$^o_2$), so we have interchanged these for comparisons. Furthermore, there is no uniformity between the two sets of results, i.e. for some transitions CIV3 f-values are larger whereas the reverse is true for LANL, and therefore, as for Sc~VI, we do not assess their results to be accurate, and subsequently do not discuss these further.

We now focus on  comparisons between our results with GRASP and those of Froese Fischer et al. \cite{mchf}  with MCHF. For most transitions listed in Table~A for Sc~VI, agreements between the two calculations are highly satisfactory, and discrepancies for a few  are within $\sim$50\% -- see for example, 2--41 and 3--41. This is particularly true for the strong transitions, as for a few weaker ones, the differences are up to two orders of magnitude -- see 1--34 and 2--32, for both of which f $\sim$ 10$^{-3}$. For weak transitions, the combined effect from different components can produce very different results because of their small magnitudes, and this depends on methods, codes, and more importantly, on CI. Such discrepancies are common among different calculations, but such transitions often do not affect the modelling of plasmas. Almost the same conclusions apply for the transitions of Mn~X in Table~B, although the discrepancies, if any, are within a factor of two, including the very weak ones. These comparisons and good agreements between two independent calculations with differing amounts of CI confirm, once again, that our reported results are accurate whereas those of  El-Maaref et al. \cite{elm1, elm2} are not. Further assessments of A-values are made below in Section~4 through the calculations of lifetimes.

Another criterion  to assess the accuracy of A-values is to compare R, the ratio of velocity and length forms of a transition. However,  this criterion is not rigorous and is only indicative of accuracy. Nevertheless, for (almost) all strong (f $\ge$ 0.1) E1 transitions R is within 20\% of unity, for all ions. Therefore, based on this and other comparisons discussed above, we can confidently state that our radiative data for a majority of (strong) transitions are accurate  to about 20\%. Unfortunately, the same cannot be said (with confidence) about the (very) weak E1 transitions, and neither about E2, M1 and M2 ones, because of their much smaller magnitudes. Additionally, very limited comparisons are possible for the  E2, M1 and M2 transitions, because neither Froese Fischer et al. \cite{mchf} nor El-Maaref et al. \cite{elm1, elm2} have reported their f- (or A-) values. However, Bi\'{e}mont and Hansen \cite{bh} have reported A-values for M1 and E2 transitions of the 3p$^4$ configuration, and therefore in Table~C we make comparisons with our results. In spite of their very small magnitudes, the agreement between the two calculations is better than 20\% for most transitions, and the only exception is 3--4 E2 ($^3$P$_0$ -- $^1$D$_2$) for which the differences are up to a factor of two, and our results are invariably higher for all ions. Further,  assessments of the A-values are made through the lifetimes, because Fischer et al. have included their contributions in the calculations of $\tau$. 

\section{Lifetimes}

The lifetimes ($\tau_j$  =  1.0/$\Sigma_{i}$A$_{ji}$) are also listed in Tables~1--4 for the ions under consideration here, and the corresponding results of Froese Fischer et al. \cite{mchf} are also included for ready comparisons. For most levels the contributions from E1 transitions dominate in the determination of $\tau$, but the inclusion of  other types, i.e. E2, M1 and M2, not only improves the accuracy but is also useful particularly for those levels for which the E1 do not connect -- note the gaps under the MCHF listings in Tables~1--4.  Although $\tau$ is a measurable quantity, no such determinations have so far been made for the four ions under consideration. So we rely on comparisons with theoretical results. El-Maaref et al. \cite{elm1, elm2} have also reported $\tau$ for some levels of Sc~VI and Mn~X. Unfortunately, their results for Sc~VI are highly inaccurate and unreliable, because discrepancies with other calculations are up to three orders of magnitude, as discussed by us \cite{sc6} and also noted above in Section~3 regarding the f-values. Since their corresponding results for Mn~X are comparatively more accurate, these are included in Table~4. 

For most levels of Sc~VI there are no (large) discrepancies between the two calculations (GRASP and MCHF) and the differences (if any) are within $\sim$20\%. However, for two levels (31 and 36) the discrepancies are  of a factor of two. These differences are a direct consequence of the corresponding differences in the A-values between the two calculations. As an example, for level 31 (3s$^2$3p$^3$($^2$P)3d~$^3$P$^o_1$) the major contributions are from 3 transitions, namely 1/2/3 to 31 with our A-values being 7.55$\times$10$^8$, 7.10$\times$10$^7$ and 2.59$\times$10$^8$~s$^{-1}$, and of Froese Fischer et al. \cite{mchf} being 3.83$\times$10$^8$, 1.45$\times$10$^8$ and 4.01$\times$10$^8$~s$^{-1}$, for the respective transitions. Since all of these contributing transitions are weak with f $<$ 10$^{-2}$, differences in $\tau$ are understandable. Similar (dis)agreements are noted in Tables~2--4 for levels of V~VIII, Cr~IX and Mn~X. On the other hand, not only the differences with the CIV3 results of El-Maaref et al. \cite{elm2} are greater (up to a factor of 5) but also for a larger number of levels of Mn~X -- see Table~4. Therefore, their results are neither accurate nor available for a majority of levels listed in the table. Furthermore, they did not include contributions from E2, M1 and M2 transitions. Finally, $\tau$ of Froese Fischer et al. are generally as accurate as our calculations, but their results are available for only a limited number of levels.  Since $\tau$ are calculated from the A-values,   our assessment of accuracy for it is  the same, i.e. $\sim$20\%, for most levels.

\section{Conclusions}

In this paper, energies for the levels of four S-like ions (Sc~VI, V VIII, Cr IX, and Mn~X) are reported. For the calculations the GRASP code has been adopted with the inclusion of CI among 4498 CSFs. However, analogous calculations have also been performed with FAC by including a  much (much) larger CI with up to 1~59~162 CSFs. These calculations only marginally improve the energies, but are highly useful for accuracy assessments, because existing results are limited to only a few levels. Based on several comparisons, with available experimental and theoretical results (compiled on the NIST website for each ion  but not exclusively included in the paper), as well as between our two calculations, our listed energies are assessed to be accurate to within about 1\%, for most levels of all ions. For brevity, energies have been listed for only a limited number of levels, which are lowest in energies, but remaining data for higher levels can be obtained from the author on request.

Radiative rates for four types of transitions, i.e. E1, E2, M1, and M2,  are also reported for all ions, whereas mostly for E1 are available in the literature so far, and that too for a limited number of transitions. Therefore, a complete set of data presented here for a larger number of levels and their transitions are expected to be highly useful for the analysis and modelling of plasmas, including astrophysical and fusion. 

For the radiative rates (and other associated parameters including lifetimes)  there are no (major) discrepancies between  our and earlier results of Froese Fischer et al. \cite{mchf}. Based on this as well the ratio of their length and velocity forms,  our A-values for significantly strong transitions are assessed to be accurate to better than  20\%. On the other hand, similar recent results of El-Maaref et al. \cite{elm1, elm2} for Sc~VI and Mn~X are found to be highly deficient, inaccurate and unreliable. 




\begin{appendix}

\def\thesection{} 

\section{Appendix A. Supplementary data}

Owing to space limitations, only parts of Tables 5-8  are presented here, but full tables are being made available as supplemental material in conjunction with the electronic publication of this work. Supplementary data associated with this article can be found, in the online version, at doi:nn.nnnn/j.adt.2019.nn.nnn.

\end{appendix}



\clearpage
\newpage
\renewcommand{\baselinestretch}{1.0}
\footnotesize
\begin{longtable}{@{\extracolsep\fill}lrlllllll@{}}
\caption{Comparison of oscillator strengths (f-values) for some transitions  of  Sc~VI. $a{\pm}b \equiv a{\times}$10$^{{\pm}b}$.} 
Transition & &  & &    \\ 
I & J & GRASP & MCHF & CIV3 & LANL   \\
\hline \\
\endfirsthead\\
\caption[]{(continued)}
Transition &  &   & &    \\  
\cline{3-6}
I & J & GRASP & MCHF & CIV3 & LANL   \\
\hline \\
\hline\\
\endhead 
     1   &   6  &  3.929$-$2  &  4.270$-$2  &  7.75$-$2  &  5.48$-$2  \\
     1   &   7  &  1.335$-$2  &  1.461$-$2  &  2.61$-$2  &  1.85$-$2  \\
     1   &  31  &  7.048$-$3  &  3.726$-$3  &  2.67$-$2  &  ---       \\
     1   &  33  &  1.766$-$2  &  1.548$-$2  &  1.48$-$1  &  1.55$-$2  \\
     1   &  34  &  3.026$-$3  &  9.569$-$5  &  1.98$-$3  &  4.29$-$4  \\
     1   &  35  &  7.400$-$3  &  6.371$-$3  &  1.36$-$2  &  1.11$-$2  \\
     1   &  36  &  1.380$-$3  &  1.040$-$3  &  ---       &  ---       \\
     1   &  37  &  3.631$-$1  &  3.139$-$1  &  2.59$-$1  &  2.90$-$1  \\ 
     1   &  38  &  7.708$-$1  &  7.916$-$1  &  5.80$-$3  &  1.10$-$0  \\
     1   &  39  &  1.611$-$1  &  1.831$-$1  &  6.71$-$4  &  3.31$-$1  \\
     1   &  41  &  8.221$-$3  &  2.139$-$2  & ---	         &	---      \\
     1   &  42  &  1.490$-$0  &  1.481$-$0  &  5.89$-$1  &  1.86$-$0  \\
     1   &  43  &  2.384$-$1  &  2.310$-$1  &  1.06$-$1  &  2.66$-$1  \\
     1   &  44  &  1.443$-$2  &  1.391$-$2  &  7.13$-$3  &  1.56$-$2  \\
     2   &   6  &  2.180$-$2  &  2.346$-$2  &  4.26$-$2  &  3.08$-$2  \\  
     2   &   7  &  1.324$-$2  &  1.417$-$2  &  2.58$-$2  &  1.84$-$2  \\
     2   &   8  &  1.765$-$2  &  1.902$-$2  &  3.47$-$2  &  2.44$-$2  \\
     2   &  29  &  3.342$-$3  &  2.224$-$3  &  2.39$-$2  &  1.22$-$4  \\
     2   &  31  &  1.127$-$3  &  2.397$-$3  &  2.65$-$2  &  5.65$-$3  \\
     2   &  32  &  2.555$-$3  &  3.635$-$4  &  2.87$-$3  &  3.07$-$3  \\
     2   &  33  &  1.747$-$2  &  4.843$-$3  &  8.18$-$2  &  7.61$-$2  \\
     2   &  34  &  5.627$-$3  &  1.374$-$2  &  9.79$-$3  &  1.59$-$2  \\
     2   &  37  &  1.879$-$1  &  2.002$-$1  &  2.59$-$1  &  2.39$-$1  \\
     2   &  38  &  3.562$-$1  &  3.551$-$1  &  3.18$-$3  &  4.61$-$2  \\ 
     2   &  39  &  3.196$-$1  &  2.747$-$1  &  6.57$-$4  &  3.96$-$3  \\
     2   &  40  &  3.367$-$1  &  3.423$-$1  &  4.33$-$4  &  4.68$-$3  \\
     2   &  41  &  3.472$-$2  &  5.522$-$2  &  ---       &  ---       \\
     2   &  43  &  1.364$-$0  &  1.363$-$0  &  1.78$-$1  &  1.72$-$1  \\
     2   &  44  &  4.313$-$1  &  4.262$-$1  &  5.30$-$1  &  5.28$-$1  \\
     2   &  45  &  1.222$-$2  &  1.517$-$2  &  ---       &  ---       \\
     3   &   7  &  5.238$-$2  &  5.600$-$2  &  1.02$-$1  &  7.35$-$2  \\
     3   &  31  &  1.245$-$2  &  ---	    &  1.05$-$1  &  4.56$-$2  \\
     3   &  32  &  1.485$-$2  &  2.141$-$2  &  1.12$-$2  &  2.21$-$2  \\
     3   &  37  &  1.326$-$1  &  1.624$-$1  &  2.56$-$1  &  2.21$-$1  \\ 
     3   &  39  &  9.526$-$1  &  8.307$-$1  &  2.57$-$3  &  1.21$-$0  \\
     3   &  41  &  1.466$-$1  &  2.258$-$1  &  ---       &  ---       \\
     3   &  44  &  1.811$-$0  &  1.813$-$0  &  7.02$-$1  &  ---       \\
     4   &   9  &  6.556$-$2  &  6.743$-$2  &  7.43$-$2  &  7.99$-$2  \\
     4   &  26  &  1.387$-$2  &  1.189$-$2  &  1.01$-$2  &  ---       \\
     4   &  37  &  1.505$-$3  &  ---	    &  ---       &  ---       \\
     4   &  38  &  3.067$-$3  &  ---	    &  ---       &  ---       \\
     4   &  39  &  5.234$-$2  &  9.114$-$2  &  ---       &  ---       \\
     4   &  41  &  5.005$-$1  &  4.744$-$1  &  ---       &  ---       \\
     4   &  43  &  3.398$-$3  &  ---	    &  ---       &  ---       \\
     4   &  44  &  1.094$-$3  &  ---	    &  ---       &  ---       \\
     4   &  45  &  8.902$-$1  &  8.541$-$1  &  7.62$-$1  &  9.58$-$1  \\
     5   &   9  &  5.657$-$3  &  7.323$-$3  &  8.97$-$2  &  2.04$-$2  \\
     5   &  39  &  9.461$-$3  &  3.910$-$4  &  ---       &  ---       \\
     5   &  41  &  1.360$-$1  &  1.139$-$1  &  ---       &  ---       \\
\\  \hline            						        				
\end{longtable}   								   					       
			      							   					       

														       
\begin{flushleft}													       
{\small
GRASP: Present results  with the {\sc grasp} code   for  4498 level calculations \\
MCHF: Calculations of Froese Fischer et al.  \cite{mchf}  with the {\sc mchf} code \\				
CIV3: Calculations of El-Maaref et al. \cite{elm1} with the {\sc civ3} code \\
LANL: Calculations of El-Maaref et al.  \cite{elm1} with the {\sc lanl} code \\
}															       
\end{flushleft} 


\clearpage
\newpage

\renewcommand{\baselinestretch}{1.0}
\footnotesize
\begin{longtable}{@{\extracolsep\fill}lrlllllll@{}}
\caption{Comparison of oscillator strengths (f-values) for some transitions  of  Mn X. $a{\pm}b \equiv a{\times}$10$^{{\pm}b}$.} 
Transition & & & &    \\ 
I & J & GRASP & MCHF & CIV3 & LANL   \\
\hline \\
\endfirsthead\\
\caption[]{(continued)}
Transition &  &   & &    \\  
\cline{3-6}
I & J & GRASP & MCHF & CIV3 & LANL   \\
\hline \\
\hline\\
\endhead 
  1  &    6   &  4.202$-$2   & 4.463$-$2  &  7.63$-$2	&  5.53$-$2  \\
  1  &    7   &  1.497$-$2   & 1.615$-$2  &  2.55$-$2	&  1.96$-$2  \\
  1  &   15   &  3.220$-$4   & 2.869$-$4  &  1.76$-$5	&  3.30$-$4  \\
  1  &   16   &  9.914$-$6   & 1.025$-$5  &  3.15$-$5	&  ---	     \\
  1  &   17   &  7.792$-$5   & 7.167$-$5  &  4.63$-$6	&  ---     \\
  1  &   28   &  1.000$-$3   & 5.096$-$4  &  8.12$-$5	& ---	     \\
  1  &   29   &  8.811$-$3   & 8.572$-$3  &  2.10$-$2	&  9.82$-$3  \\
  1  &   33   &  2.577$-$3   & 1.156$-$3  &  1.56$-$3	&  ---	     \\
  1  &   34   &  2.680$-$2   & 2.482$-$2  &  1.50$-$1	&  2.31$-$1  \\
  1  &   35   &  3.676$-$3   & 3.480$-$3  &  1.34$-$2	&  6.84$-$3  \\
  1  &   37   &  5.577$-$1   & 5.630$-$1  &  6.26$-$1	&  8.05$-$1  \\
  1  &   38   &  3.549$-$1   & 3.163$-$1  &  1.80$-$1	&  2.24$-$1  \\
  1  &   41   &  1.904$-$2   & 3.194$-$2  &  9.98$-$2	&  1.99$-$1  \\
  1  &   42   &  1.040$-$0   & 1.025$-$0  &  4.02$-$1	&  1.31$-$0  \\
  1  &   43   &  1.567$-$1   & 1.477$-$1  &  7.30$-$2	&  1.47$-$1  \\
  1  &   44   &  8.654$-$3   & 8.181$-$3  &  4.92$-$3	&  8.18$-$3  \\
  2  &    6   &  2.371$-$2   & 2.459$-$2  &  4.19$-$2	&  3.18$-$2  \\
  2  &    7   &  1.439$-$2   & 1.487$-$2  &  2.53$-$2	&  1.89$-$2  \\
  2  &    8   &  1.924$-$2   & 2.017$-$2  &  3.37$-$2	&  2.51$-$2  \\
  2  &   15   &  5.575$-$5   & 5.698$-$5  &  9.02$-$5	&  ---	     \\
  2  &   17   &  1.415$-$4   & 1.184$-$4  &  1.17$-$4	&  ---	     \\
  2  &   27   &  1.589$-$3   & 1.109$-$3  &  1.41$-$2	&  8.35$-$3  \\
  2  &   28   &  6.806$-$4   & 7.738$-$4  &  2.02$-$3	&  1.56$-$3  \\
  2  &   29   &  5.353$-$4   &  ---	         &  2.10$-$2	&  ---	     \\
  2  &   33   &  1.655$-$2   & 1.313$-$2  &  7.77$-$2	&  1.50$-$2  \\
  2  &   34   &  9.642$-$3   &  ---	         &  8.40$-$2	&  1.09$-$2  \\
  2  &   37   &  2.252$-$1   & 2.160$-$1  &  1.58$-$1	&  2.80$-$1  \\
  2  &   38   &  7.670$-$3   & 3.950$-$2  &  1.81$-$1	&  1.35$-$1  \\
  2  &   40   &  2.484$-$1   &  ---	         &  1.35$-$1	&  3.42$-$1  \\
  2  &   41   &  1.697$-$1   & 1.726$-$1  &  1.01$-$1	&  2.44$-$1  \\
  2  &   43   &  9.492$-$1   &  ---	         &  3.68$-$1	&  1.18$-$0  \\
  2  &   44   &  3.019$-$1   & 2.950$-$1  &  1.24$-$1	&  3.71$-$1  \\
  3  &    7   &  5.605$-$2   & 5.770$-$2  &  9.73$-$2	&  7.45$-$2  \\
  3  &   17   &  9.743$-$7   &  ---	         &  4.57$-$4	&  ---	     \\
  3  &   28   &  2.013$-$2   & 1.733$-$2  &  7.74$-$3	&  2.08$-$2  \\
  3  &   29   &  2.088$-$3   & 2.185$-$3  &  8.10$-$2	&  1.65$-$3  \\
  3  &   38   &  1.140$-$2   &  ---	         &  1.75$-$1	&  1.12$-$1  \\
  3  &   41   &  4.655$-$1   & 4.816$-$1  &  3.88$-$1	&  6.59$-$1  \\
  3  &   44   &  1.272$-$0   & 1.264$-$0  &  4.78$-$1	&  1.62$-$0  \\
  4  &    9   &  6.888$-$2   & 7.149$-$2  &  7.99$-$2	&  8.12$-$2  \\
  4$^c$ &26   &  9.559$-$3   & 8.716$-$3  &  8.14$-$3	& ---	     \\
  4  &   36   &  4.443$-$4   & 2.163$-$4  &  1.12$-$5	&  ---	     \\
  4$^c$ &45   &  5.963$-$1   & 5.750$-$1  &  5.44$-$1	&  6.38$-$1  \\
  4  &   46   &  1.126$-$0   &  ---	         &  9.59$-$1	&  1.48$-$0  \\
  4  &   53   &  4.347$-$3   &  ---	         &  6.62$-$3	&  1.34$-$3  \\
  5  &    9   &  1.162$-$2   & 1.328$-$2  &  1.01$-$2	&  ---	     \\
  5  &   53   &  2.063$-$0   & 2.025$-$0  &  2.16$-$0	&  2.68$-$0  \\
\\  \hline            								                	 
\end{longtable}   								   					       
			      							   					       

														       
\begin{flushleft}													       
{\small
GRASP: Present results  with the {\sc grasp} code   for  4498 level calculations \\
MCHF: Calculations of Froese Fischer et al.  \cite{mchf} with the {\sc mchf} code \\				
CIV3: Calculations of El-Maaref et al. \cite{elm2} with the {\sc civ3} code \\
LANL: Calculations of El-Maaref et al. \cite{elm2} with the {\sc lanl} code \\
c: CIV3 and LANL results of \cite{elm2} have been interchanged for the 4--26/45 transitions --  see text for details \\
}															       
\end{flushleft} 

\clearpage
\newpage

\renewcommand{\baselinestretch}{1.0}
\footnotesize
\begin{longtable}{@{\extracolsep\fill}lrlllllllllllll@{}}
\caption{Comparison of A-values (s$^{-1}$) for some M1 and E2 transitions. $a{\pm}b \equiv a{\times}$10$^{{\pm}b}$.} 
Transition &  & \multicolumn{4}{c}{Bi\'{e}mont and Hansen \cite{bh} }  & \multicolumn{4}{c}{Present Results}   \\   \\
\cline{4-11} \\
I & J & Type & Sc VI & V VIII & Cr IX & Mn X &  Sc VI & V VIII & Cr IX & Mn X  \\  \\
\hline \\
\endfirsthead\\
\caption[]{(continued)}
Transition &  &  \multicolumn{4}{c}{Bi\'{e}mont and Hansen \cite{bh} }  & \multicolumn{4}{c}{Present Results}  \\  \\
\cline{4-11} \\
I & J & Type & Sc VI & V VIII & Cr IX & Mn X &  Sc VI & V VIII & Cr IX & Mn X  \\  \\
\hline \\
\hline\\
\endhead 
  1   & 2   & M1   & 8.823$-$1   & 4.753$-$0   & 1.049$+$1   & 2.183$+$1   & 7.830$-$1   & 4.557$-$0   & 1.003$+$1   & 2.101$+$1 \\
  1   & 2   & E2   & 2.111$-$5   & 1.967$-$4   & 5.497$-$4   & 1.425$-$3   & 1.870$-$5   & 1.793$-$4   & 4.997$-$4   & 1.315$-$2 \\
  1   & 3   & E2   & 1.241$-$4   & 9.088$-$4   & 2.200$-$3   & 4.883$-$3   & 1.126$-$4   & 8.568$-$4   & 2.086$-$3   & 4.721$-$3 \\
  1   & 4   & M1   & 4.192$-$0   & 1.622$+$1   & 3.002$+$1   & 5.338$+$1   & 4.407$-$0   & 1.681$+$1   & 3.078$+$1   & 5.452$+$1 \\
  1   & 4   & E2   & 1.003$-$2   & 3.146$-$2   & 5.434$-$2   & 9.194$-$2   & 1.200$-$2   & 3.592$-$2   & 6.055$-$2   & 1.009$-$1 \\
  1   & 5   & E2   & 2.075$-$1   & 5.128$-$1   & 7.624$-$1   & 1.087$-$0   & 2.139$-$1   & 5.573$-$1   & 8.384$-$1   & 1.216$-$0 \\
  2   & 3   & M1   & 7.655$-$2   & 2.163$-$1   & 2.877$-$1   & 3.116$-$1   & 7.635$-$2   & 2.284$-$1   & 3.159$-$1   & 3.619$-$1 \\
  2   & 4   & M1   & 8.490$-$1   & 2.602$-$0   & 4.194$-$0   & 6.432$-$0   & 9.462$-$1   & 2.884$-$0   & 4.627$-$0   & 7.085$-$0 \\
  2   & 4   & E2   & 6.314$-$4   & 1.325$-$3   & 1.805$-$3   & 2.368$-$3   & 8.786$-$4   & 1.765$-$3   & 2.362$-$3   & 3.058$-$3 \\
  2   & 5   & M1   & 4.972$+$1   & 1.855$+$2   & 3.355$+$2   & 5.806$+$2   & 5.117$+$1   & 1.910$+$2   & 3.436$-$2   & 5.952$+$2 \\
  3   & 4   & E2   & 1.405$-$4   & 3.754$-$4   & 5.884$-$4   & 9.092$-$4   & 2.516$-$4   & 5.795$-$4   & 8.560$-$4   & 1.256$-$3 \\
  4   & 5   & E2   & 4.250$-$0   & 5.519$-$0   & 6.291$-$0   & 7.184$-$0   & 4.504$-$0   & 6.128$-$0   & 7.059$-$0   & 8.126$-$0 \\
\\  \hline            								                	 
\end{longtable}   								   					       
			      							   					       

														       
\begin{flushleft}													       
{\small
}															       
\end{flushleft} 

\clearpage
\newpage


\TableExplanation

\bigskip
\renewcommand{\arraystretch}{1.0}

\section*{Table 1.\label{tbl1te} Energies (Ryd) for the lowest 61 levels of Sc~VI and their lifetimes ($\tau$, s).}
\begin{tabular}{@{}p{1in}p{6in}@{}}
Index            & Level Index \\
Configuration    & The configuration to which the level belongs \\
Level             & The $LSJ$ designation of the level \\
NIST              & Energies compiled by NIST and available at the website  {\tt http://www.nist.gov/pml/data/asd.cfm} \\
GRASP          & Present energies from the {\sc grasp} code  with 41  configurations and 4498 level calculations \\
FAC             & Present energies from the {\sc fac} code  with 1~59~162 level calculations \\
MCHF          & Earlier calculations of Froese Fischer et al.  \cite{mchf} with the {\sc mchf} code \\ 
$\tau$ (MCHF)       & Lifetime (in s) of the level  from the MCHF calculations of  Froese Fischer et al.  \cite{mchf}  \\
$\tau$ (GRASP)       & Lifetime (in s) of the level  from present calculations with the {\sc grasp} code \\

\end{tabular}
\label{ExplTable1}

\bigskip
\renewcommand{\arraystretch}{1.0}

\section*{Table 2.\label{tbl2te} Energies (Ryd) for the lowest 110 levels of V~VIII and their lifetimes ($\tau$, s).}
\begin{tabular}{@{}p{1in}p{6in}@{}}
Index            & Level Index \\
Configuration    & The configuration to which the level belongs \\
Level             & The $LSJ$ designation of the level \\
NIST              & Energies compiled by NIST and available at the website  {\tt http://www.nist.gov/pml/data/asd.cfm} \\
GRASP          & Present energies from the {\sc grasp} code  with 41  configurations and 4498 level calculations \\
FAC             & Present energies from the {\sc fac} code  with 1~59~162 level calculations \\
MCHF          & Earlier calculations of Froese Fischer et al.  \cite{mchf} with the {\sc mchf} code \\ 
$\tau$ (MCHF)       & Lifetime (in s) of the level  from the MCHF calculations of  Froese Fischer et al.  \cite{mchf}  \\
$\tau$ (GRASP)       & Lifetime (in s) of the level  from present calculations with the {\sc grasp} code \\

\end{tabular}
\label{ExplTable2}

\bigskip
\renewcommand{\arraystretch}{1.0}

\section*{Table 3.\label{tbl3te} Energies (Ryd) for the lowest  143 levels of Cr~IX and their lifetimes ($\tau$, s). }
\begin{tabular}{@{}p{1in}p{6in}@{}}
Index            & Level Index \\
Configuration    & The configuration to which the level belongs \\
Level             & The $LSJ$ designation of the level \\
NIST              & Energies compiled by NIST and available at the website  {\tt http://www.nist.gov/pml/data/asd.cfm} \\
GRASP          & Present energies from the {\sc grasp} code  with 41  configurations and 4498 level calculations \\
FAC             & Present energies from the {\sc fac} code  with 1~59~162 level calculations \\
MCHF          & Earlier calculations of Froese Fischer et al.  \cite{mchf} with the {\sc mchf} code \\ 
$\tau$ (MCHF)       & Lifetime (in s) of the level  from the MCHF calculations of  Froese Fischer et al.  \cite{mchf}  \\
$\tau$ (GRASP)       & Lifetime (in s) of the level  from present calculations with the {\sc grasp} code \\

\end{tabular}
\label{ExplTable3}

\bigskip
\renewcommand{\arraystretch}{1.0}

\section*{Table 4.\label{tbl4te} Energies (Ryd) for the lowest 190 levels of Mn~X and their lifetimes ($\tau$, s). }
\begin{tabular}{@{}p{1in}p{6in}@{}}
Index            & Level Index \\
Configuration    & The configuration to which the level belongs \\
Level             & The $LSJ$ designation of the level \\
NIST              & Energies compiled by NIST and available at the website  {\tt http://www.nist.gov/pml/data/asd.cfm} \\
GRASP          & Present energies from the {\sc grasp} code  with 41  configurations and 4498 level calculations \\
FAC             & Present energies from the {\sc fac} code  with 1~59~162 level calculations \\
MCHF          & Earlier calculations of Froese Fischer et al.  \cite{mchf} with the {\sc mchf} code \\ 
$\tau$ (MCHF)       & Lifetime (in s) of the level  from the MCHF calculations of  Froese Fischer et al.  \cite{mchf}  \\
$\tau$ (GRASP)       & Lifetime (in s) of the level  from present calculations with the {\sc grasp} code \\

\end{tabular}
\label{ExplTable4}

\bigskip
\renewcommand{\arraystretch}{1.0}


\bigskip
\section*{Table 5.\label{tbl5e}  Transition wavelengths ($\lambda_{ij}$ in $\rm \AA$), radiative rates (A$_{ji}$ in s$^{-1}$),
 oscillator strengths (f$_{ij}$, dimensionless), and line strengths (S, in atomic units) for electric dipole (E1), and 
A$_{ji}$ for electric quadrupole (E2), magnetic dipole (M1), and magnetic quadrupole (M2) transitions of Sc~VI.
The ratio R(E1) of velocity and length forms of A-values for E1 transitions is listed in the last column.}
\begin{tabular}{@{}p{1in}p{6in}@{}}
$i$ and $j$         & The lower ($i$) and upper ($j$) levels of a transition as defined in Table 1.\\
$\lambda_{ij}$      & Transition wavelength (in ${\rm \AA}$) \\
A$^{E1}_{ji}$       & Radiative transition probability (in s$^{-1}$) for the E1 transitions \\
f$^{E1}_{ij}$       & Absorption oscillator strength (dimensionless) for the E1 transitions \\
S$^{E1}$            & Line strength in atomic unit (a.u.), 1 a.u. = 6.460$\times$10$^{-36}$ cm$^2$ esu$^2$ for the E1 transitions \\
A$^{E2}_{ji}$       & Radiative transition probability (in s$^{-1}$) for the E2 transitions \\
A$^{M1}_{ji}$       & Radiative transition probability (in s$^{-1}$) for the M1 transitions \\
A$^{M2}_{ji}$       & Radiative transition probability (in s$^{-1}$) for the M2 transitions \\
R(E1)                     & Ratio of velocity and length forms of A- (or f- and S-) values for the E1 transitions \\
$a{\pm}b$ &  $\equiv a\times{10^{{\pm}b}}$ \\
\end{tabular}
\label{ExplTable5}

\bigskip
\section*{Table 6.\label{tbl6te}  Transition wavelengths ($\lambda_{ij}$ in $\rm \AA$), radiative rates (A$_{ji}$ in s$^{-1}$),
 oscillator strengths (f$_{ij}$, dimensionless), and line strengths (S, in atomic units) for electric dipole (E1), and 
A$_{ji}$ for electric quadrupole (E2), magnetic dipole (M1), and magnetic quadrupole (M2) transitions of V~VIII.
The ratio R(E1) of velocity and length forms of A-values for E1 transitions is listed in the last column.}
\begin{tabular}{@{}p{1in}p{6in}@{}}
$i$ and $j$         & The lower ($i$) and upper ($j$) levels of a transition as defined in Table 2.\\
$\lambda_{ij}$      & Transition wavelength (in ${\rm \AA}$) \\
A$^{E1}_{ji}$       & Radiative transition probability (in s$^{-1}$) for the E1 transitions \\
f$^{E1}_{ij}$       & Absorption oscillator strength (dimensionless) for the E1 transitions \\
S$^{E1}$            & Line strength in atomic unit (a.u.), 1 a.u. = 6.460$\times$10$^{-36}$ cm$^2$ esu$^2$ for the E1 transitions \\
A$^{E2}_{ji}$       & Radiative transition probability (in s$^{-1}$) for the E2 transitions \\
A$^{M1}_{ji}$       & Radiative transition probability (in s$^{-1}$) for the M1 transitions \\
A$^{M2}_{ji}$       & Radiative transition probability (in s$^{-1}$) for the M2 transitions \\
R(E1)                     & Ratio of velocity and length forms of A- (or f- and S-) values for the E1 transitions \\
$a{\pm}b$ &  $\equiv a\times{10^{{\pm}b}}$ \\
\end{tabular}
\label{ExplTable6}

\bigskip
\section*{Table 7.\label{tbl7te}  Transition wavelengths ($\lambda_{ij}$ in $\rm \AA$), radiative rates (A$_{ji}$ in s$^{-1}$),
 oscillator strengths (f$_{ij}$, dimensionless), and line strengths (S, in atomic units) for electric dipole (E1), and 
A$_{ji}$ for electric quadrupole (E2), magnetic dipole (M1), and magnetic quadrupole (M2) transitions of Cr~IX.
 The ratio R(E1) of velocity and length forms of A-values for E1 transitions is listed in the last column.}
\begin{tabular}{@{}p{1in}p{6in}@{}}
$i$ and $j$         & The lower ($i$) and upper ($j$) levels of a transition as defined in Table 3.\\
$\lambda_{ij}$      & Transition wavelength (in ${\rm \AA}$) \\
A$^{E1}_{ji}$       & Radiative transition probability (in s$^{-1}$) for the E1 transitions \\
f$^{E1}_{ij}$       & Absorption oscillator strength (dimensionless) for the E1 transitions \\
S$^{E1}$            & Line strength in atomic unit (a.u.), 1 a.u. = 6.460$\times$10$^{-36}$ cm$^2$ esu$^2$ for the E1 transitions \\
A$^{E2}_{ji}$       & Radiative transition probability (in s$^{-1}$) for the E2 transitions \\
A$^{M1}_{ji}$       & Radiative transition probability (in s$^{-1}$) for the M1 transitions \\
A$^{M2}_{ji}$       & Radiative transition probability (in s$^{-1}$) for the M2 transitions \\
R(E1)                     & Ratio of velocity and length forms of A- (or f- and S-) values for the E1 transitions \\
$a{\pm}b$ &  $\equiv a\times{10^{{\pm}b}}$ \\
\end{tabular}
\label{ExplTable7}

\bigskip
\section*{Table 8.\label{tbl8te}  Transition wavelengths ($\lambda_{ij}$ in $\rm \AA$), radiative rates (A$_{ji}$ in s$^{-1}$),
 oscillator strengths (f$_{ij}$, dimensionless), and line strengths (S, in atomic units) for electric dipole (E1), and 
A$_{ji}$ for electric quadrupole (E2), magnetic dipole (M1), and magnetic quadrupole (M2) transitions of Mn~X.
 The ratio R(E1) of velocity and length forms of A-values for E1 transitions is listed in the last column.}
\begin{tabular}{@{}p{1in}p{6in}@{}}
$i$ and $j$         & The lower ($i$) and upper ($j$) levels of a transition as defined in Table 4.\\
$\lambda_{ij}$      & Transition wavelength (in ${\rm \AA}$) \\
A$^{E1}_{ji}$       & Radiative transition probability (in s$^{-1}$) for the E1 transitions \\
f$^{E1}_{ij}$       & Absorption oscillator strength (dimensionless) for the E1 transitions \\
S$^{E1}$            & Line strength in atomic unit (a.u.), 1 a.u. = 6.460$\times$10$^{-36}$ cm$^2$ esu$^2$ for the E1 transitions \\
A$^{E2}_{ji}$       & Radiative transition probability (in s$^{-1}$) for the E2 transitions \\
A$^{M1}_{ji}$       & Radiative transition probability (in s$^{-1}$) for the M1 transitions \\
A$^{M2}_{ji}$       & Radiative transition probability (in s$^{-1}$) for the M2 transitions \\
R(E1)                     & Ratio of velocity and length forms of A- (or f- and S-) values for the E1 transitions \\
$a{\pm}b$ &  $\equiv a\times{10^{{\pm}b}}$ \\
\end{tabular}
\label{ExplTable8}


\datatables 



\setlength{\LTleft}{0pt}
\setlength{\LTright}{0pt} 


\setlength{\tabcolsep}{0.5\tabcolsep}

\renewcommand{\arraystretch}{1.0}

\footnotesize 

\begin{longtable}{@{\extracolsep\fill}rlllrrrrrrr@{}}
\caption{Energies (Ryd) for  the lowest 61 levels of Sc~VI and their lifetimes ($\tau$, s).   $a{\pm}b \equiv a{\times}$10$^{{\pm}b}$. See page\ \pageref{tbl1te} for Explanation of Tables. Levels with $\star$ are interchanged between the 3p$^3$($^2$P)3d and 3p$^3$($^2$D)3d configurations of the NIST listings -- see text in section 2.}
Index  &     Configuration        & Level                & NIST  &  GRASP        &    FAC & MCHF   & CIV3 &  $\tau$ (GRASP) &   $\tau$ (MCHF)  \\  \\
\hline\\
\endfirsthead\\
\caption[]{(continued)}
Index  &     Configuration        & Level                & NIST  &  GRASP        &    FAC & MCHF   & CIV3 &   $\tau$ (GRASP) & $\tau$ (MCHF)  \\  \\
\hline\\
\endhead
    1 & 3s$^2$3p$^4$			&  $^3$P$  _{2}$  & 0.0000	  &  0.0000 &  0.0000 & 0.0000 & 0.0000   &		&	      \\
    2 & 3s$^2$3p$^4$			&  $^3$P$  _{1}$  & 0.0305	  &  0.0299 &  0.0296 & 0.0293 & 0.0302   & 1.277$-$00  &	      \\
    3 & 3s$^2$3p$^4$			&  $^3$P$  _{0}$  & 0.0406	  &  0.0401 &  0.0397 & 0.0387 & 0.0410   & 1.308$+$01  &	      \\
    4 & 3s$^2$3p$^4$			&  $^1$D$  _{2}$  & 0.1949	  &  0.2124 &  0.2103 & 0.2010 & 0.1910   & 1.863$-$01  &	      \\
    5 & 3s$^2$3p$^4$			&  $^1$S$  _{0}$  & 0.4486	  &  0.4703 &  0.4647 & 0.4511 & 0.4542   & 1.789$-$02  &	      \\
    6 & 3s3p$^5$			&  $^3$P$^o_{2}$  & 1.5979	  &  1.5907 &  1.5830 & 1.5828 & 1.5907   & 9.440$-$10  & 8.729$-$10  \\
    7 & 3s3p$^5$			&  $^3$P$^o_{1}$  & 1.6239	  &  1.6163 &  1.6084 & 1.6074 & 1.6185   & 9.226$-$10  & 8.630$-$10  \\
    8 & 3s3p$^5$			&  $^3$P$^o_{0}$  & 1.6383	  &  1.6304 &  1.6224 & 1.6210 & 1.6325   & 9.179$-$10  & 8.614$-$10  \\
    9 & 3s3p$^5$			&  $^1$P$^o_{1}$  & 2.0455	  &  2.0857 &  2.0598 & 2.0400 & 2.0473   & 3.178$-$10  & 3.198$-$10  \\
   10 & 3s$^2$3p$^3$($^4$S)3d		&  $^5$D$^o_{0}$  & 		  &  2.1904 &  2.1584 & 2.1698 & 2.1261   & 9.769$-$08  & 1.090$-$07  \\
   11 & 3s$^2$3p$^3$($^4$S)3d		&  $^5$D$^o_{1}$  & 		  &  2.1907 &  2.1586 & 2.1702 & 2.1278   & 1.131$-$07  & 1.262$-$07  \\
   12 & 3s$^2$3p$^3$($^4$S)3d		&  $^5$D$^o_{2}$  & 		  &  2.1911 &  2.1590 & 2.1709 & 2.1291   & 1.696$-$07  & 1.922$-$07  \\
   13 & 3s$^2$3p$^3$($^4$S)3d		&  $^5$D$^o_{3}$  & 		  &  2.1918 &  2.1596 & 2.1718 & 2.1300   & 4.208$-$07  & 5.099$-$07  \\
   14 & 3s$^2$3p$^3$($^4$S)3d		&  $^5$D$^o_{4}$  & 		  &  2.1929 &  2.1607 &        & 2.1305   & 1.356$-$01  &	      \\
   15 & 3s$^2$3p$^3$($^2$D)3d		&  $^3$D$^o_{2}$  & 		  &  2.3761 &  2.3374 & 2.3348 & 2.3418   & 7.903$-$08  & 1.022$-$08  \\
   16 & 3s$^2$3p$^3$($^2$D)3d		&  $^3$D$^o_{3}$  & 		  &  2.3768 &  2.3380 & 2.3357 & 2.3395   & 3.896$-$07  & 5.137$-$07  \\
   17 & 3s$^2$3p$^3$($^2$D)3d		&  $^3$D$^o_{1}$  & 		  &  2.3806 &  2.3419 & 2.3393 & 2.3432   & 6.961$-$08  & 9.028$-$08  \\
   18 & 3s$^2$3p$^3$($^2$D)3d		&  $^1$S$^o_{0}$  & 		  &  2.4412 &  2.3971 & 2.3985 & 2.4395   & 1.152$-$07  & 1.500$-$07  \\
   19 & 3s$^2$3p$^3$($^2$D)3d		&  $^3$F$^o_{2}$  & 		  &  2.4449 &  2.4060 & 2.3983 & 2.4023   & 6.152$-$07  & 7.105$-$07  \\
   20 & 3s$^2$3p$^3$($^2$D)3d		&  $^3$F$^o_{3}$  & 		  &  2.4541 &  2.4150 & 2.4073 & 2.4086   & 2.484$-$07  & 3.148$-$07  \\
   \hline  											      
			      							   					       

\\
\end{longtable}

\begin{longtable}{@{\extracolsep\fill}rllrrrrrrrrr@{}}
\caption{Energies (Ryd) for  the lowest 110 levels of V~VIII and their lifetimes ($\tau$, s).   $a{\pm}b \equiv a{\times}$10$^{{\pm}b}$. See page\ \pageref{tbl2te} for Explanation of Tables.}
Index  &     Configuration        & Level                & NIST  &  GRASP        &    FAC & MCHF   &   $\tau$ (GRASP) &   $\tau$ (MCHF)  \\  \\
\hline\\
\endfirsthead\\
\caption[]{(continued)}
Index  &     Configuration        & Level                & NIST  &  GRASP        &    FAC & MCHF   &   $\tau$ (GRASP) &   $\tau$ (MCHF)  \\  \\
\hline\\
\endhead
    1   &   3s$^2$3p$^4$		         &  $^3$P$  _{2}$      & 0.0000   &  0.0000  &  0.0000  & 0.0000   &		   &		   \\
    2   &   3s$^2$3p$^4$		         &  $^3$P$  _{1}$      & 0.0547   &  0.0539  &  0.0536  & 0.0535   &   2.194$-$01  &		   \\
    3   &   3s$^2$3p$^4$		         &  $^3$P$  _{0}$      & 0.0691   &  0.0687  &  0.0683  & 0.0667   &   4.362$-$00  &		   \\
    4   &   3s$^2$3p$^4$		         &  $^1$D$  _{2}$      & 0.2467   &  0.2629  &  0.2612  & 0.2534   &   5.067$-$02  &		   \\
    5   &   3s$^2$3p$^4$		         &  $^1$S$  _{0}$      & 0.5526   &  0.5764  &  0.5703  & 0.5552   &   5.058$-$03  &		   \\
    6   &   3s3p$^5$			         &  $^3$P$^o_{2}$      & 1.9819   &  1.9802  &  1.9702  & 1.9669   &   5.684$-$10  & 5.417$-$10    \\
    7   &   3s3p$^5$			         &  $^3$P$^o_{1}$      & 2.0267   &  2.0246  &  2.0143  & 2.0100   &   5.496$-$10  & 5.257$-$10    \\
    8   &   3s3p$^5$			         &  $^3$P$^o_{0}$      & 2.0526   &  2.0500  &  2.0395  & 2.0346   &   5.477$-$10  & 5.261$-$10    \\
    9   &   3s3p$^5$			         &  $^1$P$^o_{1}$      & 2.5350   &  2.5814  &  2.5512  & 2.5313   &   1.921$-$10  & 1.921$-$10    \\
   10   &   3s$^2$3p$^3$($^4$S)3d	         &  $^5$D$^o_{0}$      &          &  2.7409  &  2.7029  & 2.7169   &   3.377$-$08  & 3.632$-$08    \\
   11   &   3s$^2$3p$^3$($^4$S)3d	         &  $^5$D$^o_{1}$      &          &  2.7415  &  2.7035  & 2.7178   &   3.847$-$08  & 4.109$-$08    \\
   12   &   3s$^2$3p$^3$($^4$S)3d	         &  $^5$D$^o_{2}$      &          &  2.7425  &  2.7045  & 2.7192   &   5.846$-$08  & 6.272$-$08    \\
   13   &   3s$^2$3p$^3$($^4$S)3d	         &  $^5$D$^o_{3}$      &          &  2.7440  &  2.7059  & 2.7211   &   1.535$-$07  & 1.716$-$07    \\
   14   &   3s$^2$3p$^3$($^4$S)3d	         &  $^5$D$^o_{4}$      &          &  2.7469  &  2.7087  & 	   &   6.416$-$02  &	           \\
   15   &   3s$^2$3p$^3$($^2$D)3d	         &  $^3$D$^o_{2}$      &          &  2.9578  &  2.9133  & 2.9139   &   5.155$-$08  & 6.336$-$08    \\
   16   &   3s$^2$3p$^3$($^2$D)3d	         &  $^3$D$^o_{3}$      &          &  2.9618  &  2.9172  & 2.9184   &   3.933$-$06  & 3.852$-$06    \\
   17   &   3s$^2$3p$^3$($^2$D)3d	         &  $^3$D$^o_{1}$      &          &  2.9690  &  2.9243  & 2.9251   &   6.139$-$08  & 7.799$-$08    \\
   18   &   3s$^2$3p$^3$($^2$D)3d	         &  $^3$F$^o_{2}$      &          &  3.0327  &  2.9880  & 2.9844   &   2.402$-$07  & 2.664$-$07    \\
   19   &   3s$^2$3p$^3$($^2$D)3d	         &  $^1$S$^o_{0}$      &          &  3.0387  &  2.9888  & 2.9950   &   4.434$-$08  & 5.505$-$08    \\
   20   &   3s$^2$3p$^3$($^2$D)3d	         &  $^3$F$^o_{3}$      &          &  3.0490  &  3.0040  & 3.0005   &   1.138$-$07  & 1.364$-$07    \\
   \hline  											      
			      							   					       

\\
\end{longtable}
\clearpage
\newpage

\begin{longtable}{@{\extracolsep\fill}rllrrrrrrrrr@{}}
\caption{Energies (Ryd) for  the lowest 143 levels of Cr~IX and their lifetimes ($\tau$, s).   $a{\pm}b \equiv a{\times}$10$^{{\pm}b}$. See page\ \pageref{tbl3te} for Explanation of Tables.}
Index  &     Configuration        & Level                & NIST  &  GRASP        &    FAC & MCHF   &   $\tau$ (GRASP) &   $\tau$ (MCHF)  \\  \\
\hline\\
\endfirsthead\\
\caption[]{(continued)}
Index  &     Configuration        & Level                & NIST  &  GRASP        &    FAC & MCHF  &   $\tau$ (GRASP) &   $\tau$ (MCHF)  \\  \\
\hline\\
\endhead
    1   &   3s$^2$3p$^4$			&   $^3$P$  _{2}$    & 0.0000	&  0.0000  & 0.0000  &  0.0000    &		   &  		   \\
    2   &   3s$^2$3p$^4$			&   $^3$P$  _{1}$    & 0.0713	&  0.0702  & 0.0699  &  0.0702    &   9.971$-$02   &  		   \\
    3   &   3s$^2$3p$^4$			&   $^3$P$  _{0}$    & 0.0870	&  0.0868  & 0.0863  &  0.0845    &   3.145$-$00   &  		   \\
    4   &   3s$^2$3p$^4$			&   $^1$D$  _{2}$    & 0.2760	&  0.2916  & 0.2900  &  0.2832    &   2.819$-$02   &  		   \\
    5   &   3s$^2$3p$^4$			&   $^1$S$  _{0}$    & 0.6092	&  0.6336  & 0.6274  &  0.6119    &   2.845$-$03   &  		   \\
    6   &   3s3p$^5$				&   $^3$P$^o_{2}$    & 2.1786	&  2.1791  & 2.1680  &  2.1639    &   4.658$-$10   &  4.462$-$10   \\
    7   &   3s3p$^5$				&   $^3$P$^o_{1}$    & 2.2355	&  2.2358  & 2.2243  &  2.2191    &   4.473$-$10   &  4.299$-$10   \\
    8   &   3s3p$^5$				&   $^3$P$^o_{0}$    & 2.2692	&  2.2688  & 2.2573  &  2.2514    &   4.467$-$10   &  4.314$-$10   \\
    9   &   3s3p$^5$				&   $^1$P$^o_{1}$    & 2.7845	&  2.8330  & 2.8010  &  2.7817    &   1.586$-$10   &  1.581$-$10   \\
   10   &   3s$^2$3p$^3$($^4$S)3d		&   $^5$D$^o_{0}$    &       	&  3.0134  & 2.9730  &  2.9885    &   2.127$-$08   &  2.253$-$08   \\
   11   &   3s$^2$3p$^3$($^4$S)3d		&   $^5$D$^o_{1}$    &       	&  3.0143  & 2.9739  &  2.9897    &   2.389$-$08   &  2.503$-$08   \\
   12   &   3s$^2$3p$^3$($^4$S)3d		&   $^5$D$^o_{2}$    &       	&  3.0156  & 2.9752  &  2.9916    &   3.646$-$08   &  3.827$-$08   \\
   13   &   3s$^2$3p$^3$($^4$S)3d		&   $^5$D$^o_{3}$    &       	&  3.0176  & 2.9771  &  2.9941    &   9.764$-$08   &  1.059$-$07   \\
   14   &   3s$^2$3p$^3$($^4$S)3d		&   $^5$D$^o_{4}$    &       	&  3.0218  & 2.9813  &            &   4.686$-$02   &       	   \\
   15   &   3s$^2$3p$^3$($^2$D)3d		&   $^3$D$^o_{2}$    &       	&  3.2444  & 3.1977  &  3.2001    &   3.524$-$08   &  4.175$-$08   \\
   16   &   3s$^2$3p$^3$($^2$D)3d		&   $^3$D$^o_{3}$    &       	&  3.2519  & 3.2050  &  3.2085    &   1.791$-$06   &  1.568$-$06   \\
   17   &   3s$^2$3p$^3$($^2$D)3d		&   $^3$D$^o_{1}$    &       	&  3.2614  & 3.2143  &  3.2173    &   4.721$-$08   &  5.770$-$08   \\
   18   &   3s$^2$3p$^3$($^2$D)3d		&   $^3$F$^o_{2}$    &       	&  3.3239  & 3.2767  &  3.2758    &   1.417$-$07   &  1.534$-$07   \\
   19   &   3s$^2$3p$^3$($^2$D)3d		&   $^1$S$^o_{0}$    &       	&  3.3358  & 3.2835  &  3.2925    &   2.969$-$08   &  3.263$-$08   \\
   20   &   3s$^2$3p$^3$($^2$D)3d		&   $^3$F$^o_{3}$    &       	&  3.3446  & 3.2971  &  3.2962    &   7.952$-$08   &  9.268$-$08   \\
\hline  											      
			      							   					       

\\
\end{longtable}

\begin{longtable}{@{\extracolsep\fill}rllrrrrrrrr@{}}
\caption{Energies (Ryd) for  the lowest 190 levels of Mn~X and their lifetimes ($\tau$, s).  $a{\pm}b \equiv a{\times}$10$^{{\pm}b}$. See page\ \pageref{tbl4te} for Explanation of Tables.}
Index  &     Configuration        & Level                & NIST  &  GRASP        &    FAC & CIV3   &   $\tau$ (GRASP) &  $\tau$ (CIV3) & $\tau$ (MCHF)  \\  \\
\hline\\
\endfirsthead\\
\caption[]{(continued)}
Index  &     Configuration        & Level                & NIST  &  GRASP        &    FAC & CIV3   &   $\tau$ (GRASP) &  $\tau$ (CIV3) & $\tau$ (MCHF)  \\  \\
\hline\\
\endhead
    1  &  3s$^2$3p$^4$  		    &	$^3$P$  _{2}$	  & 0.0000   & 0.0000	& 0.0000 & 0.0000 &		 &	     	&  	     	\\
    2  &  3s$^2$3p$^4$  		    &	$^3$P$  _{1}$	  & 0.0913   & 0.0901	& 0.0898 & 0.0897 & 4.759$-$02   &	     	&  	     	\\
    3  &  3s$^2$3p$^4$  		    &	$^3$P$  _{0}$	  & 0.1075   & 0.1075	& 0.1070 & 0.1095 & 2.728$-$00   &	     	&  	     	\\
    4  &  3s$^2$3p$^4$  		    &	$^1$D$  _{2}$	  & 0.3082   & 0.3233	& 0.3218 & 0.3028 & 1.620$-$02   &	     	&  	     	\\
    5  &  3s$^2$3p$^4$  		    &	$^1$S$  _{0}$	  & 0.6702   & 0.6950	& 0.6887 & 0.6799 & 1.654$-$03   &	     	&  	     	\\
    6  &  3s3p$^5$			    &	$^3$P$^o_{2}$	  & 2.3791   & 2.3816	& 2.3695 & 2.3777 & 3.910$-$10   &  2.21$-$10	&  3.757$-$10	\\
    7  &  3s3p$^5$			    &	$^3$P$^o_{1}$	  & 2.4502   & 2.4525	& 2.4400 & 2.4526 & 3.725$-$10   &  2.17$-$10	&  3.588$-$10	\\
    8  &  3s3p$^5$			    &	$^3$P$^o_{0}$	  & 2.4934   & 2.4949	& 2.4824 & 2.4894 & 3.730$-$10   &  2.14$-$10	&  3.615$-$10	\\
    9  &  3s3p$^5$			    &	$^1$P$^o_{1}$	  & 3.0382   & 3.0886	& 3.0550 & 3.0634 & 1.345$-$10   &  1.03$-$10	&  1.336$-$10	\\
   10  &  3s$^2$3p$^3$($^4$S)3d 	    &	$^5$D$^o_{0}$	  & 	     & 3.2858	& 3.2433 & 3.2200 & 1.389$-$08   &	     	&  1.450$-$08	\\
   11  &  3s$^2$3p$^3$($^4$S)3d 	    &	$^5$D$^o_{1}$	  & 	     & 3.2870	& 3.2445 & 3.2251 & 1.530$-$08   &	     	&  1.573$-$08	\\
   12  &  3s$^2$3p$^3$($^4$S)3d 	    &	$^5$D$^o_{2}$	  & 	     & 3.2887	& 3.2462 & 3.2288 & 2.348$-$08   &	     	&  2.413$-$08	\\
   13  &  3s$^2$3p$^3$($^4$S)3d 	    &	$^5$D$^o_{3}$	  & 	     & 3.2913	& 3.2487 & 3.2310 & 6.380$-$08   &	     	&  6.734$-$08	\\
   14  &  3s$^2$3p$^3$($^4$S)3d 	    &	$^5$D$^o_{4}$	  & 	     & 3.2972	& 3.2546 & 3.2325 & 3.517$-$02   &	     	&  	     	\\
   15  &  3s$^2$3p$^3$($^2$D)3d 	    &	$^3$D$^o_{2}$	  & 	     & 3.5295	& 3.4809 & 3.5096 & 2.401$-$08   &  1.46$-$08	&  2.751$-$08	\\
   16  &  3s$^2$3p$^3$($^2$D)3d 	    &	$^3$D$^o_{3}$	  & 	     & 3.5425	& 3.4935 & 3.5162 & 8.182$-$07   &  4.52$-$07	&  7.973$-$07	\\
   17  &  3s$^2$3p$^3$($^2$D)3d 	    &	$^3$D$^o_{1}$	  & 	     & 3.5546	& 3.5055 & 3.5198 & 3.441$-$08   &  3.84$-$08	&  4.041$-$08	\\
   18  &  3s$^2$3p$^3$($^2$D)3d 	    &	$^3$F$^o_{2}$	  & 	     & 3.6154	& 3.5661 & 3.5809 & 8.114$-$08   &	     	&  8.595$-$08	\\
   19  &  3s$^2$3p$^3$($^2$D)3d 	    &	$^1$S$^o_{0}$	  & 	     & 3.6335	& 3.5792 & 3.6492 & 2.080$-$08   &	     	&  2.505$-$08	\\
   20  &  3s$^2$3p$^3$($^2$D)3d 	    &	$^3$F$^o_{3}$	  & 	     & 3.6408	& 3.5912 & 	  & 5.682$-$08   &	     	&  6.461$-$08	\\
\hline  											      
			      							   					       

\\
\end{longtable}

\clearpage
\newpage

\renewcommand{\arraystretch}{1.0}

\footnotesize 

\begin{longtable}{@{\extracolsep\fill}rrrrrrrrrr@{}}
\caption{Transition wavelengths ($\lambda_{ij}$ in $\rm \AA$), radiative rates (A$_{ji}$ in s$^{-1}$), oscillator strengths (f$_{ij}$, dimensionless), and line     
strengths (S, in atomic units) for electric dipole (E1), and A$_{ji}$ for E2, M1, and M2 transitions in Sc~VI.  The last column gives R(E1), the ratio of velocity and length forms of A-values for E1 transitions. $a{\pm}b \equiv a{\times}$10$^{{\pm}b}$. See page\  \pageref{tbl5te} for Explanation of Tables.}
$i$ & $j$ & $\lambda_{ij}$ & A$^{{\rm E1}}_{ji}$  & f$^{{\rm E1}}_{ij}$ & S$^{{\rm E1}}$ & A$^{{\rm E2}}_{ji}$  & A$^{{\rm M1}}_{ji}$ & A$^{{\rm M2}}_{ji}$ & R(E1)   \\ \\
\hline\\
\endfirsthead\\
\caption[]{(continued)}
$i$ & $j$ & $\lambda_{ij}$ & A$^{{\rm E1}}_{ji}$  & f$^{{\rm E1}}_{ij}$ & S$^{{\rm E1}}$ & A$^{{\rm E2}}_{ji}$  & A$^{{\rm M1}}_{ji}$ & A$^{{\rm M2}}_{ji}$  & R(E1)  \\ \\
\hline\\
\endhead
    1 &    2 &  3.053$+$04 &  0.000$+$00 &  0.000$+$00 &  0.000$+$00 &  1.870$-$05 &  7.830$-$01 &  0.000$+$00 &   0.0$+$00 \\
    1 &    3 &  2.272$+$04 &  0.000$+$00 &  0.000$+$00 &  0.000$+$00 &  1.126$-$04 &  0.000$+$00 &  0.000$+$00 &   0.0$+$00 \\
    1 &    4 &  4.291$+$03 &  0.000$+$00 &  0.000$+$00 &  0.000$+$00 &  1.200$-$02 &  4.407$+$00 &  0.000$+$00 &   0.0$+$00 \\
    1 &    5 &  1.938$+$03 &  0.000$+$00 &  0.000$+$00 &  0.000$+$00 &  2.139$-$01 &  0.000$+$00 &  0.000$+$00 &   0.0$+$00 \\
    1 &    6 &  5.729$+$02 &  7.985$+$08 &  3.929$-$02 &  3.705$-$01 &  0.000$+$00 &  0.000$+$00 &  2.817$+$00 &   9.0$-$01 \\
    1 &    7 &  5.638$+$02 &  4.670$+$08 &  1.335$-$02 &  1.239$-$01 &  0.000$+$00 &  0.000$+$00 &  2.381$-$02 &   9.1$-$01 \\
    1 &    8 &  5.589$+$02 &  0.000$+$00 &  0.000$+$00 &  0.000$+$00 &  0.000$+$00 &  0.000$+$00 &  1.568$+$00 &   0.0$+$00 \\
    1 &    9 &  4.369$+$02 &  2.548$+$07 &  4.376$-$04 &  3.147$-$03 &  0.000$+$00 &  0.000$+$00 &  8.702$+$00 &   7.0$-$01 \\
    1 &   10 &  4.160$+$02 &  0.000$+$00 &  0.000$+$00 &  0.000$+$00 &  0.000$+$00 &  0.000$+$00 &  1.467$-$01 &   0.0$+$00 \\
    1 &   11 &  4.160$+$02 &  4.315$+$06 &  6.717$-$05 &  4.599$-$04 &  0.000$+$00 &  0.000$+$00 &  5.934$-$04 &   9.0$-$01 \\
    1 &   12 &  4.159$+$02 &  5.855$+$06 &  1.518$-$04 &  1.039$-$03 &  0.000$+$00 &  0.000$+$00 &  1.707$-$01 &   9.1$-$01 \\
    1 &   13 &  4.158$+$02 &  2.373$+$06 &  8.609$-$05 &  5.892$-$04 &  0.000$+$00 &  0.000$+$00 &  4.890$-$03 &   9.2$-$01 \\
    1 &   14 &  4.155$+$02 &  0.000$+$00 &  0.000$+$00 &  0.000$+$00 &  0.000$+$00 &  0.000$+$00 &  7.345$+$00 &   0.0$+$00 \\
    1 &   15 &  3.835$+$02 &  1.078$+$07 &  2.377$-$04 &  1.501$-$03 &  0.000$+$00 &  0.000$+$00 &  4.037$+$00 &   1.3$+$00 \\
    1 &   16 &  3.834$+$02 &  2.494$+$06 &  7.696$-$05 &  4.857$-$04 &  0.000$+$00 &  0.000$+$00 &  8.302$-$01 &   2.9$+$00 \\
    1 &   17 &  3.828$+$02 &  2.708$+$06 &  3.570$-$05 &  2.249$-$04 &  0.000$+$00 &  0.000$+$00 &  4.383$+$00 &   1.1$+$00 \\
    1 &   18 &  3.733$+$02 &  0.000$+$00 &  0.000$+$00 &  0.000$+$00 &  0.000$+$00 &  0.000$+$00 &  1.391$+$01 &   0.0$+$00 \\
    1 &   19 &  3.727$+$02 &  7.746$+$05 &  1.613$-$05 &  9.898$-$05 &  0.000$+$00 &  0.000$+$00 &  9.465$-$02 &   1.1$+$00 \\
    1 &   20 &  3.713$+$02 &  1.831$+$06 &  5.300$-$05 &  3.239$-$04 &  0.000$+$00 &  0.000$+$00 &  5.130$-$01 &   1.2$+$00 \\
\hline


\\
\end{longtable}

\renewcommand{\arraystretch}{1.0}

\footnotesize 

\begin{longtable}{@{\extracolsep\fill}rrrrrrrrrr@{}}
\caption{Transition wavelengths ($\lambda_{ij}$ in $\rm \AA$), radiative rates (A$_{ji}$ in s$^{-1}$), oscillator strengths (f$_{ij}$, dimensionless), and line     
strengths (S, in atomic units) for electric dipole (E1), and A$_{ji}$ for E2, M1, and M2 transitions in V~VIII.  The last column gives R(E1), the ratio of velocity and length forms of A-values for E1 transitions. $a{\pm}b \equiv a{\times}$10$^{{\pm}b}$. See page\ \pageref{tbl6te} for Explanation of Tables.}
$i$ & $j$ & $\lambda_{ij}$ & A$^{{\rm E1}}_{ji}$  & f$^{{\rm E1}}_{ij}$ & S$^{{\rm E1}}$ & A$^{{\rm E2}}_{ji}$  & A$^{{\rm M1}}_{ji}$ & A$^{{\rm M2}}_{ji}$ & R(E1)   \\ \\
\hline\\
\endfirsthead\\
\caption[]{(continued)}
$i$ & $j$ & $\lambda_{ij}$ & A$^{{\rm E1}}_{ji}$  & f$^{{\rm E1}}_{ij}$ & S$^{{\rm E1}}$ & A$^{{\rm E2}}_{ji}$  & A$^{{\rm M1}}_{ji}$ & A$^{{\rm M2}}_{ji}$  & R(E1)  \\ \\
\hline\\
\endhead
    1 &    2 &  1.692$+$04 &  0.000$+$00 &  0.000$+$00 &  0.000$+$00 &  1.793$-$04 &  4.557$+$00 &  0.000$+$00 &   0.0$+$00 \\
    1 &    3 &  1.326$+$04 &  0.000$+$00 &  0.000$+$00 &  0.000$+$00 &  8.568$-$04 &  0.000$+$00 &  0.000$+$00 &   0.0$+$00 \\
    1 &    4 &  3.466$+$03 &  0.000$+$00 &  0.000$+$00 &  0.000$+$00 &  3.592$-$02 &  1.681$+$01 &  0.000$+$00 &   0.0$+$00 \\
    1 &    5 &  1.581$+$03 &  0.000$+$00 &  0.000$+$00 &  0.000$+$00 &  5.572$-$01 &  0.000$+$00 &  0.000$+$00 &   0.0$+$00 \\
    1 &    6 &  4.602$+$02 &  1.323$+$09 &  4.201$-$02 &  3.183$-$01 &  0.000$+$00 &  0.000$+$00 &  6.187$+$00 &   9.2$-$01 \\
    1 &    7 &  4.501$+$02 &  7.978$+$08 &  1.454$-$02 &  1.077$-$01 &  0.000$+$00 &  0.000$+$00 &  1.213$-$01 &   9.2$-$01 \\
    1 &    8 &  4.445$+$02 &  0.000$+$00 &  0.000$+$00 &  0.000$+$00 &  0.000$+$00 &  0.000$+$00 &  3.122$+$00 &   0.0$+$00 \\
    1 &    9 &  3.530$+$02 &  8.108$+$07 &  9.089$-$04 &  5.282$-$03 &  0.000$+$00 &  0.000$+$00 &  1.728$+$01 &   7.6$-$01 \\
    1 &   10 &  3.325$+$02 &  0.000$+$00 &  0.000$+$00 &  0.000$+$00 &  0.000$+$00 &  0.000$+$00 &  2.757$-$01 &   0.0$+$00 \\
    1 &   11 &  3.324$+$02 &  1.321$+$07 &  1.313$-$04 &  7.182$-$04 &  0.000$+$00 &  0.000$+$00 &  3.344$-$03 &   9.2$-$01 \\
    1 &   12 &  3.323$+$02 &  1.702$+$07 &  2.817$-$04 &  1.541$-$03 &  0.000$+$00 &  0.000$+$00 &  3.125$-$01 &   9.4$-$01 \\
    1 &   13 &  3.321$+$02 &  6.493$+$06 &  1.503$-$04 &  8.216$-$04 &  0.000$+$00 &  0.000$+$00 &  2.636$-$02 &   9.5$-$01 \\
    1 &   14 &  3.318$+$02 &  0.000$+$00 &  0.000$+$00 &  0.000$+$00 &  0.000$+$00 &  0.000$+$00 &  1.549$+$01 &   0.0$+$00 \\
    1 &   15 &  3.081$+$02 &  1.713$+$07 &  2.438$-$04 &  1.236$-$03 &  0.000$+$00 &  0.000$+$00 &  9.143$+$00 &   1.2$+$00 \\
    1 &   16 &  3.077$+$02 &  7.492$+$03 &  1.489$-$07 &  7.539$-$07 &  0.000$+$00 &  0.000$+$00 &  2.112$+$00 &   1.7$+$02 \\
    1 &   17 &  3.069$+$02 &  5.619$+$06 &  4.762$-$05 &  2.406$-$04 &  0.000$+$00 &  0.000$+$00 &  8.811$+$00 &   1.1$+$00 \\
    1 &   18 &  3.005$+$02 &  2.939$+$06 &  3.978$-$05 &  1.968$-$04 &  0.000$+$00 &  0.000$+$00 &  4.084$-$01 &   1.1$+$00 \\
    1 &   19 &  2.999$+$02 &  0.000$+$00 &  0.000$+$00 &  0.000$+$00 &  0.000$+$00 &  0.000$+$00 &  2.526$+$01 &   0.0$+$00 \\
    1 &   20 &  2.989$+$02 &  3.916$+$06 &  7.342$-$05 &  3.612$-$04 &  0.000$+$00 &  0.000$+$00 &  1.150$+$00 &   1.2$+$00 \\
\hline


\\
\end{longtable}

\clearpage
\newpage

\renewcommand{\arraystretch}{1.0}

\footnotesize 

\begin{longtable}{@{\extracolsep\fill}rrrrrrrrrr@{}}
\caption{Transition wavelengths ($\lambda_{ij}$ in $\rm \AA$), radiative rates (A$_{ji}$ in s$^{-1}$), oscillator strengths (f$_{ij}$, dimensionless), and line     
strengths (S, in atomic units) for electric dipole (E1), and A$_{ji}$ for E2, M1, and M2 transitions in Cr~IX.  The last column gives R(E1), the ratio of velocity and length forms of A-values for E1 transitions. $a{\pm}b \equiv a{\times}$10$^{{\pm}b}$. See page\ \pageref{tbl7te} for Explanation of Tables.}
$i$ & $j$ & $\lambda_{ij}$ & A$^{{\rm E1}}_{ji}$  & f$^{{\rm E1}}_{ij}$ & S$^{{\rm E1}}$ & A$^{{\rm E2}}_{ji}$  & A$^{{\rm M1}}_{ji}$ & A$^{{\rm M2}}_{ji}$  & R(E1)  \\ \\
\hline\\
\endfirsthead\\
\caption[]{(continued)}
$i$ & $j$ & $\lambda_{ij}$ & A$^{{\rm E1}}_{ji}$  & f$^{{\rm E1}}_{ij}$ & S$^{{\rm E1}}$ & A$^{{\rm E2}}_{ji}$  & A$^{{\rm M1}}_{ji}$ & A$^{{\rm M2}}_{ji}$ & R(E1)   \\ \\
\hline\\
\endhead
    1 &    2 &  1.298$+$04 &  0.000$+$00 &  0.000$+$00 &  0.000$+$00 &  4.997$-$04 &  1.003$+$01 &  0.000$+$00 &   0.0$+$00 \\
    1 &    3 &  1.050$+$04 &  0.000$+$00 &  0.000$+$00 &  0.000$+$00 &  2.086$-$03 &  0.000$+$00 &  0.000$+$00 &   0.0$+$00 \\
    1 &    4 &  3.125$+$03 &  0.000$+$00 &  0.000$+$00 &  0.000$+$00 &  6.055$-$02 &  3.078$+$01 &  0.000$+$00 &   0.0$+$00 \\
    1 &    5 &  1.438$+$03 &  0.000$+$00 &  0.000$+$00 &  0.000$+$00 &  8.384$-$01 &  0.000$+$00 &  0.000$+$00 &   0.0$+$00 \\
    1 &    6 &  4.182$+$02 &  1.612$+$09 &  4.225$-$02 &  2.908$-$01 &  0.000$+$00 &  0.000$+$00 &  8.744$+$00 &   9.2$-$01 \\
    1 &    7 &  4.076$+$02 &  9.911$+$08 &  1.481$-$02 &  9.937$-$02 &  0.000$+$00 &  0.000$+$00 &  2.443$-$01 &   9.3$-$01 \\
    1 &    8 &  4.016$+$02 &  0.000$+$00 &  0.000$+$00 &  0.000$+$00 &  0.000$+$00 &  0.000$+$00 &  4.180$+$00 &   0.0$+$00 \\
    1 &    9 &  3.217$+$02 &  1.293$+$08 &  1.203$-$03 &  6.371$-$03 &  0.000$+$00 &  0.000$+$00 &  2.308$+$01 &   7.8$-$01 \\
    1 &   10 &  3.024$+$02 &  0.000$+$00 &  0.000$+$00 &  0.000$+$00 &  0.000$+$00 &  0.000$+$00 &  3.532$-$01 &   0.0$+$00 \\
    1 &   11 &  3.023$+$02 &  2.178$+$07 &  1.791$-$04 &  8.912$-$04 &  0.000$+$00 &  0.000$+$00 &  7.855$-$03 &   9.3$-$01 \\
    1 &   12 &  3.022$+$02 &  2.727$+$07 &  3.734$-$04 &  1.857$-$03 &  0.000$+$00 &  0.000$+$00 &  3.848$-$01 &   9.4$-$01 \\
    1 &   13 &  3.020$+$02 &  1.020$+$07 &  1.952$-$04 &  9.703$-$04 &  0.000$+$00 &  0.000$+$00 &  5.562$-$02 &   9.6$-$01 \\
    1 &   14 &  3.016$+$02 &  0.000$+$00 &  0.000$+$00 &  0.000$+$00 &  0.000$+$00 &  0.000$+$00 &  2.117$+$01 &   0.0$+$00 \\
    1 &   15 &  2.809$+$02 &  2.339$+$07 &  2.766$-$04 &  1.279$-$03 &  0.000$+$00 &  0.000$+$00 &  1.287$+$01 &   1.2$+$00 \\
    1 &   16 &  2.802$+$02 &  1.882$+$05 &  3.103$-$06 &  1.431$-$05 &  0.000$+$00 &  0.000$+$00 &  3.175$+$00 &   1.6$+$00 \\
    1 &   17 &  2.794$+$02 &  8.568$+$06 &  6.017$-$05 &  2.767$-$04 &  0.000$+$00 &  0.000$+$00 &  1.179$+$01 &   1.0$+$00 \\
    1 &   18 &  2.742$+$02 &  5.781$+$06 &  6.515$-$05 &  2.940$-$04 &  0.000$+$00 &  0.000$+$00 &  8.460$-$01 &   1.0$+$00 \\
    1 &   19 &  2.732$+$02 &  0.000$+$00 &  0.000$+$00 &  0.000$+$00 &  0.000$+$00 &  0.000$+$00 &  3.185$+$01 &   0.0$+$00 \\
    1 &   20 &  2.725$+$02 &  5.689$+$06 &  8.864$-$05 &  3.975$-$04 &  0.000$+$00 &  0.000$+$00 &  1.602$+$00 &   1.2$+$00 \\
\hline


\\
\end{longtable}

\renewcommand{\arraystretch}{1.0}

\footnotesize 

\begin{longtable}{@{\extracolsep\fill}rrrrrrrrrr@{}}
\caption{Transition wavelengths ($\lambda_{ij}$ in $\rm \AA$), radiative rates (A$_{ji}$ in s$^{-1}$), oscillator strengths (f$_{ij}$, dimensionless), and line     
strengths (S, in atomic units) for electric dipole (E1), and A$_{ji}$ for E2, M1, and M2 transitions in Mn~X.  The last column gives R(E1), the ratio of velocity and length forms of A-values for E1 transitions. $a{\pm}b \equiv a{\times}$10$^{{\pm}b}$. See page\ \pageref{tbl8te} for Explanation of Tables.}
$i$ & $j$ & $\lambda_{ij}$ & A$^{{\rm E1}}_{ji}$  & f$^{{\rm E1}}_{ij}$ & S$^{{\rm E1}}$ & A$^{{\rm E2}}_{ji}$  & A$^{{\rm M1}}_{ji}$ & A$^{{\rm M2}}_{ji}$  & R(E1)  \\ \\
\hline\\
\endfirsthead\\
\caption[]{(continued)}
$i$ & $j$ & $\lambda_{ij}$ & A$^{{\rm E1}}_{ji}$  & f$^{{\rm E1}}_{ij}$ & S$^{{\rm E1}}$ & A$^{{\rm E2}}_{ji}$  & A$^{{\rm M1}}_{ji}$ & A$^{{\rm M2}}_{ji}$  & R(E1)  \\ \\
\hline\\
\endhead
    1 &    2 &  1.012$+$04 &  0.000$+$00 &  0.000$+$00 &  0.000$+$00 &  1.315$-$03 &  2.101$+$01 &  0.000$+$00 &   0.0$+$00 \\
    1 &    3 &  8.478$+$03 &  0.000$+$00 &  0.000$+$00 &  0.000$+$00 &  4.721$-$03 &  0.000$+$00 &  0.000$+$00 &   0.0$+$00 \\
    1 &    4 &  2.818$+$03 &  0.000$+$00 &  0.000$+$00 &  0.000$+$00 &  1.009$-$01 &  5.452$+$01 &  0.000$+$00 &   0.0$+$00 \\
    1 &    5 &  1.311$+$03 &  0.000$+$00 &  0.000$+$00 &  0.000$+$00 &  1.216$+$00 &  0.000$+$00 &  0.000$+$00 &   0.0$+$00 \\
    1 &    6 &  3.826$+$02 &  1.914$+$09 &  4.202$-$02 &  2.646$-$01 &  0.000$+$00 &  0.000$+$00 &  1.207$+$01 &   9.2$-$01 \\
    1 &    7 &  3.716$+$02 &  1.206$+$09 &  1.497$-$02 &  9.156$-$02 &  0.000$+$00 &  0.000$+$00 &  4.639$-$01 &   9.3$-$01 \\
    1 &    8 &  3.652$+$02 &  0.000$+$00 &  0.000$+$00 &  0.000$+$00 &  0.000$+$00 &  0.000$+$00 &  5.451$+$00 &   0.0$+$00 \\
    1 &    9 &  2.950$+$02 &  1.947$+$08 &  1.524$-$03 &  7.402$-$03 &  0.000$+$00 &  0.000$+$00 &  3.000$+$01 &   7.9$-$01 \\
    1 &   10 &  2.773$+$02 &  0.000$+$00 &  0.000$+$00 &  0.000$+$00 &  0.000$+$00 &  0.000$+$00 &  4.386$-$01 &   0.0$+$00 \\
    1 &   11 &  2.772$+$02 &  3.492$+$07 &  2.414$-$04 &  1.102$-$03 &  0.000$+$00 &  0.000$+$00 &  1.804$-$02 &   9.3$-$01 \\
    1 &   12 &  2.771$+$02 &  4.232$+$07 &  4.871$-$04 &  2.222$-$03 &  0.000$+$00 &  0.000$+$00 &  4.463$-$01 &   9.5$-$01 \\
    1 &   13 &  2.769$+$02 &  1.559$+$07 &  2.509$-$04 &  1.143$-$03 &  0.000$+$00 &  0.000$+$00 &  1.117$-$01 &   9.7$-$01 \\
    1 &   14 &  2.764$+$02 &  0.000$+$00 &  0.000$+$00 &  0.000$+$00 &  0.000$+$00 &  0.000$+$00 &  2.818$+$01 &   0.0$+$00 \\
    1 &   15 &  2.582$+$02 &  3.222$+$07 &  3.220$-$04 &  1.368$-$03 &  0.000$+$00 &  0.000$+$00 &  1.746$+$01 &   1.1$+$00 \\
    1 &   16 &  2.572$+$02 &  7.138$+$05 &  9.914$-$06 &  4.198$-$05 &  0.000$+$00 &  0.000$+$00 &  4.633$+$00 &   9.5$-$03 \\
    1 &   17 &  2.564$+$02 &  1.318$+$07 &  7.792$-$05 &  3.288$-$04 &  0.000$+$00 &  0.000$+$00 &  1.539$+$01 &   1.0$+$00 \\
    1 &   18 &  2.521$+$02 &  1.115$+$07 &  1.062$-$04 &  4.405$-$04 &  0.000$+$00 &  0.000$+$00 &  1.692$+$00 &   1.0$+$00 \\
    1 &   19 &  2.508$+$02 &  0.000$+$00 &  0.000$+$00 &  0.000$+$00 &  0.000$+$00 &  0.000$+$00 &  3.883$+$01 &   0.0$+$00 \\
    1 &   20 &  2.503$+$02 &  8.182$+$06 &  1.076$-$04 &  4.433$-$04 &  0.000$+$00 &  0.000$+$00 &  2.143$+$00 &   1.2$+$00 \\
\hline


\\
\end{longtable}

\end{document}